\documentclass[journal]{IEEEtran}
\usepackage{graphicx}
\begin{document}
\title{FPGA Online Tracking Algorithm for the PANDA Straw Tube Tracker}




\author{Yutie~Liang, Hua~Ye, Martin~J~Galuska, Thomas~Gessler, Wolfgang~K\"{u}hn,~\IEEEmembership{Member,~IEEE}, Jens~S\"{o}ren~Lange, Milan~N~Wagner, Zhen'an~Liu,~\IEEEmembership{Senior Member,~IEEE}, Jingzhou~Zhao,~\IEEEmembership{Member,~IEEE,}
\thanks{Manuscript received June 24, 2016. This work was supported in part by BMBF (05P12RGFPF), the
LOEWE-Zentrum HICforFAIR and the JCHP FFE(COSY-099)
}
\thanks{Y.~T.~Liang, M.~J.~Galuska, W.~K\"{u}hn, J.~S.~Lange, and M.~N.~Wagner are with the II. Physikalisches Institut, University of Giessen, Heinrich-Buff-Ring 14, 35392, Giessen, Germany (e-mail: yutie.liang@physik.uni-giessen.de; Martin.J.Galuska@physik.uni-giessen.de; Wolfgang.Kuehn@exp2.physik.uni-giessen.de; Soeren.Lange@exp2. physik.uni-giessen.de; milan.n.wagner@physik.uni-giessen.de).}
\thanks{Z.~Liu and J.~Zhao are with the Institute of High Energy Physics, Chinese Academy of Sciences, 100049 Beijing, China (e-mail: liuza@ihep.ac.cn; zhaojz@ihep.ac.cn).}
\thanks{T.~Gessler was with the II. Physikalisches Institut, University of Giessen, Heinrich-Buff-Ring 14, 35392, Giessen, Germany, and now is with the High Energy Accelerator Research Organization (KEK), Tsukuba, Ibaraki 305-0801, Japan.(e-mail: tgessler@post.kek.jp).}
\thanks{Y.~Hua is with the Institute of High Energy Physics, Chinese Academy of Sciences.(e-mail: yehua@mail.ihep.ac.cn).
}}

\maketitle
\thispagestyle{empty}

\begin{abstract}

\textbf{A novel FPGA based online tracking algorithm for helix track reconstruction in a solenoidal field, developed for the PANDA spectrometer, is described. Employing the Straw Tube Tracker detector with 4636 straw tubes, the algorithm includes a complex track finder, and a track fitter. Implemented in VHDL, the algorithm is tested on a Xilinx Virtex-4 FX60 FPGA chip with different types of events, at different event rates. A processing time of 7 $\mu$s per event for an average of 6 charged tracks is obtained. The momentum resolution is about 3\% (4\%) for $p_t$ ($p_z$) at 1 GeV/c. Comparing to the algorithm running on a CPU chip (single core Intel Xeon E5520 at 2.26 GHz), an improvement of 3 orders of magnitude in processing time is obtained. The algorithm can handle severe overlapping of events which are typical for interaction rates above 10 MHz.}

\end{abstract}

\section{Introduction}

The PANDA (anti-Proton ANnihilations at DArmstadt) experiment \cite{PANDA} is a fixed target experiment to be constructed at the Facility for Antiproton and Ion Research (FAIR) in Darmstadt, Germany. PANDA operates with high-precision antiproton beams in the momentum range between 1.5 GeV/c and 15 GeV/c hitting a fixed target. The PANDA detector consists of a target spectrometer and a forward spectrometer. The target spectrometer will operate in a 2T solenoidal field. To provide a measurement of charged particle trajectories, a silicon Micro Vertex Detector (MVD) \cite{TDRMVD} and a Straw Tube Tracker (STT) \cite{TDRSTT} are employed.

The experiment will be operating with very high interaction rates up to average of 20 MHz in a freely streaming mode. Without any hardware trigger, up to 200 GB of raw data are streaming into the data acquisition (DAQ) system. Before mass storage, the data rate has to be reduced by at least three orders of magnitude. This will be achieved by filtering the events based on their physics content using information from reconstruction algorithms implemented on Field Programmable Gate Arrays (FPGA) as a first level and on a farm of servers with attached GPUs as a second level. To achieve reduction factors of the order of 1000, a fast online tracking algorithm running on FPGAs is used. 

FPGAs as a platform to run tracking algorithms have been employed in many experiments. In the D$\O$ experiment at  Fermilab~\cite{DPHI}, the central track trigger logic is implemented in FPGAs which compare hit inputs to a predefined list of track equations. In the CDF experiment~\cite{CDF}, the online tracking system performs pattern recognition with Associative Memories and a linear track fitter implemented in FPGAs. In the CMS experiment~\cite{CMS}, the resistive plate chambers based technical trigger system implements a tracking algorithm in FPGAs, detecting cosmic muon tracks. In the ATLAS experiment~\cite{ATLAS}, the Fast Tracker system as a part of the ATLAS trigger upgrade programme performs global tracking at 100 kHz. Here, FPGA devices are employed for track fitting calculations. 

In this paper, a sophisticated FPGA based tracking algorithm for PANDA is presented. Employing the full STT detector with 4636 straw tubes, the algorithm is designed with a complex track finder, followed by a  track fitter. Due to the high event rate of 20 MHz, serious overlapping of events has to be treated properly. The algorithm features a high tracking efficiency, low latency and sufficient  momentum resolution for event filtering in realtime.

The tracking algorithm uses the hit information from the STT~\cite{TDRSTT}. The STT is the main tracking detector for charged particles in the target spectrometer and consists of 4636 single straw tubes. Straws are gas-filled cylindrical tubes with a conductive inner layer as cathode and an anode wire stretched along the cylinder axis. An electric field between the wire and the outer conductor separates electrons and positive ions produced by a charged particle along its trajectory through the gas volume. 

By measuring the drift time of the earliest arriving electrons, information about the minimum particle track distance from the wire is obtained. The isochrone contains all space points belonging to the same electron drift time and describes a cylinder around the wire axis. The characteristic relation between drift time and isochrone is given by the electron drift velocity, which will be obtained by calibration. As a slow response detector, the electron drift time could extend up to 200 ns, depending on the distance of the electron to the wire. More details on the STT design can be found at Ref.\cite{TDRSTT}

\begin{figure}[ht!]
\centering
\includegraphics[width=3.5in]{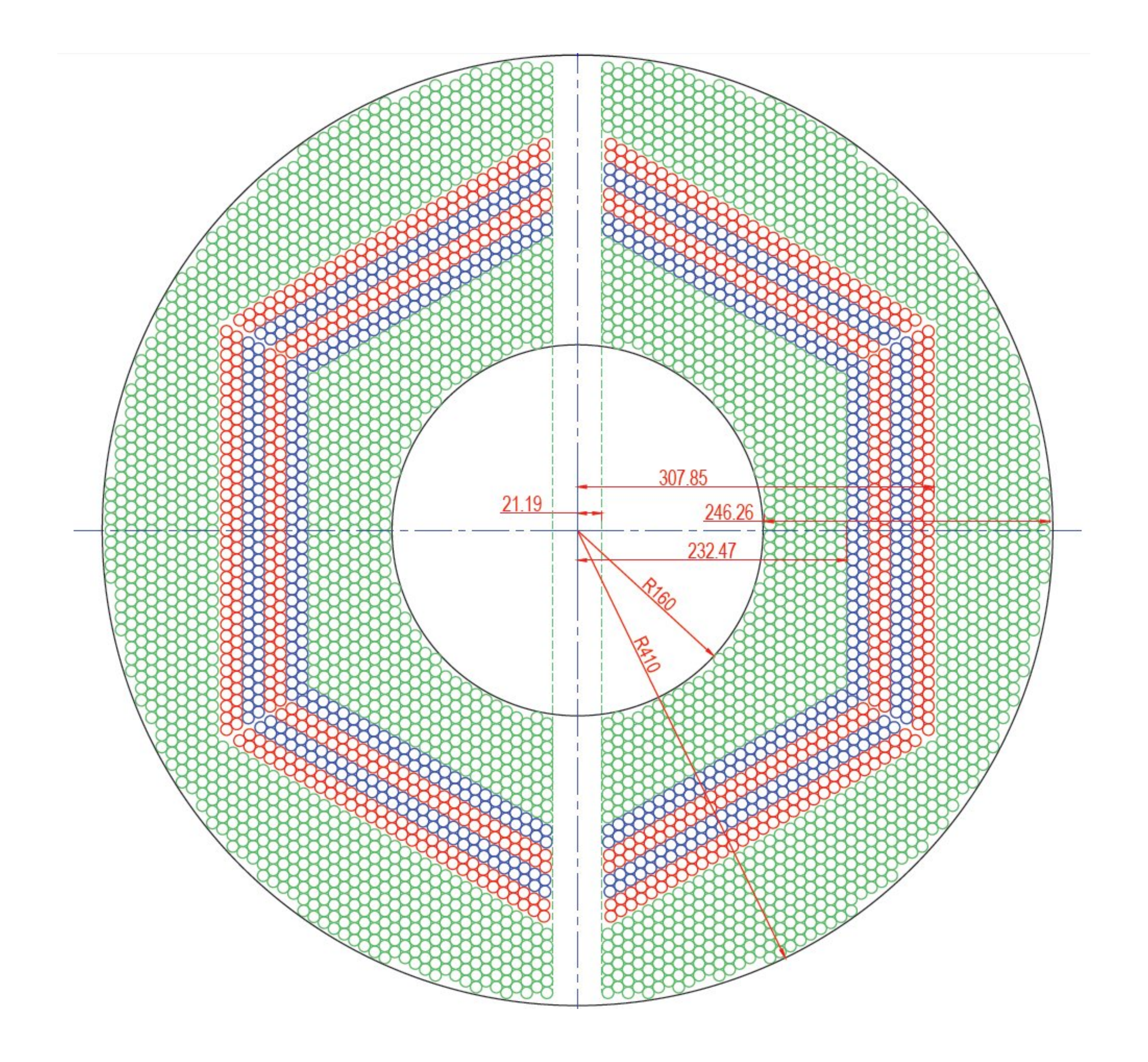}
\caption{The layout of the STT in X-Y view. The straws marked in green are parallel to the beam axis. The blue and red straws form stereo layers. \label{stt_layout}}
\end{figure}

The layout of STT is shown in Fig.~\ref{stt_layout}. The STT is divided into two semi-cylindrical volumes, each of which is filled by three sectors of straw tubes aligned in the z-direction and arranged in stacks of planar multi-layer modules. In each of the six hexagonal sectors, starting from the inner radius there are 8 straw layers parallel to the beam axis, followed by a block of 4 stereo double-layers, alternately skewed by $\pm2.9^o$ relative to the axially aligned straw layers, and again a block of 4 layers parallel to the beam axis. Furthermore, there are 7 layers parallel to the beam with a decreasing number of straws per layer to achieve the outer cylindrical shape of the STT. The inner cylindrical shape is achieved by placing a few axially aligned straws in the inner corner region of each hexagon sector. All straws have the same inner diameter of 10 mm and a length of 1500 mm, except for a few outer straws in the border region of each skewed layer, which have reduced lengths.


\section{Tracking algorithm}

Operating in a freely streaming mode, the quasi-continuous beam delivered to PANDA is organised in bursts of 2000 ns duration, with a 400 ns gap between two bursts. Assuming that a the event start time $T_0$ is provided by a fast external detector, a collection of hits from $T_0$ to $T_0+200$ ns in a burst is packed as one STT burst event, potentially containing many overlapping proton-antiproton interactions. The drift time of each hit in this burst event is approximated by the difference of the measured arrival time of the hit and  $T_0$, with the flight time (a few ns) of the particle from the interaction point to the tube being ignored. 

At 20 MHz interaction rate, an average of 40 events are expected in one burst. Due to the large electron drift time (up to 200 ns), the STT hits from the events in one burst have serious overlapping with respect to the arrival time. One obvious feature of the overlapping of events is the increase of the number of hits in a burst event, which then introduces a number of fake tracks due to the overlapping events.

With the drift time information, the drift circles of hits in the burst event are calculated and fed to the tracking algorithm which is divided into two stages, the track finder and the track fitter.

\subsection{Track finder}

\begin{figure}[ht!]
\centering
\includegraphics[width=3.5in]{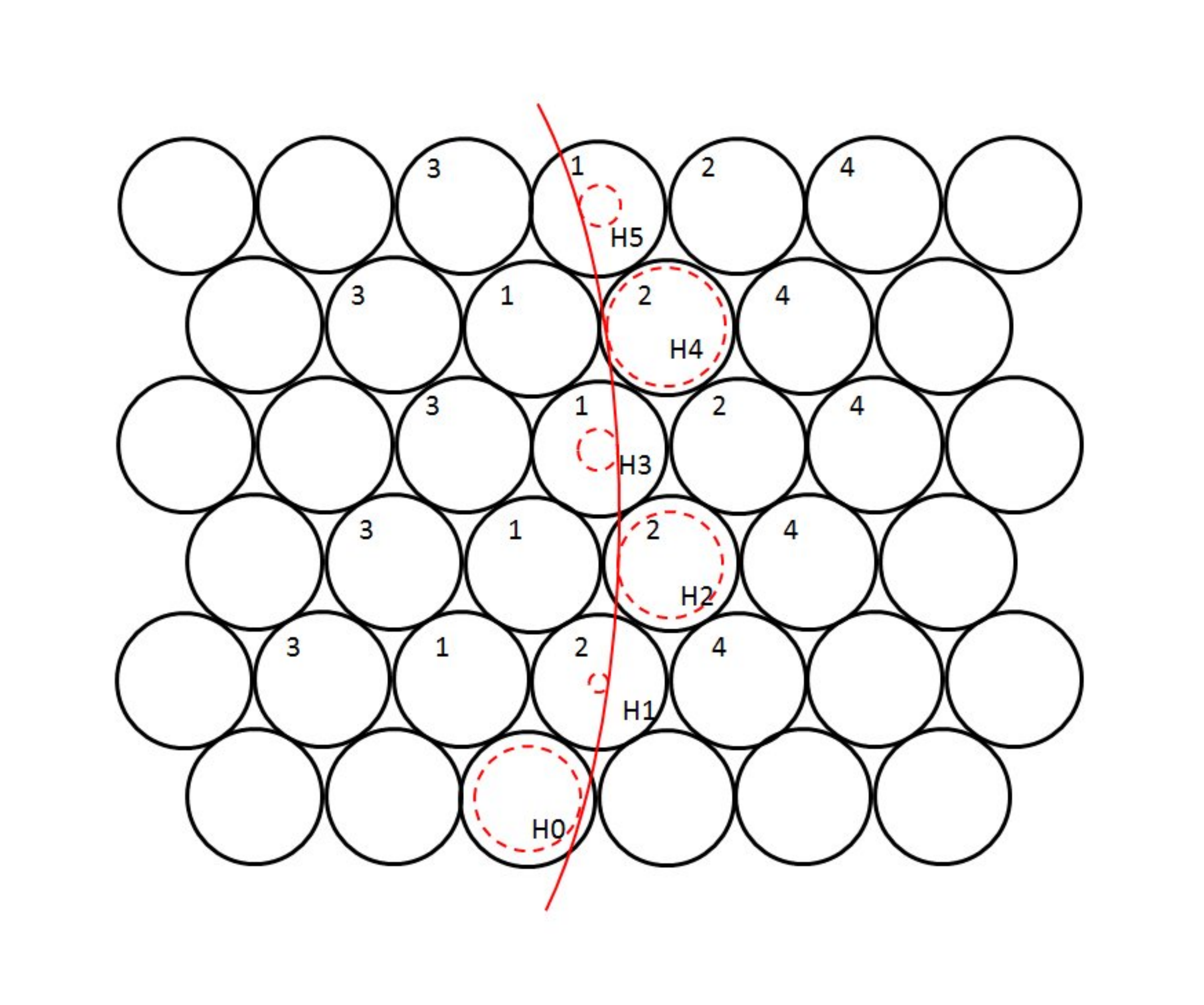}
\caption{Schematic view of the track finder. The black solid circles represent straw tubes. The red solid curve indicates the trajectory of a charged particle. The red dashed circles inside tubes represent the isochrones. \label{track_finder}}
\end{figure}

The procedure of track finding is depicted in Fig.~\ref{track_finder}. In this figure, the black solid circles represent straw tubes. The red solid curve indicates the trajectory of a charged particle. In the tubes on the trajectory, the drift circles, calculated using $T_0$ are shown as red dashed circles. 

The track finder starts from hits in the innermost layer, marked as H0.  The hits in H0 are considered as a seed and are used to create a tracklet.  Moving to the next layer, two adjacent tubes, tube 1 and 2 in the second layer are searched and the active tube is attached to the tracklet. If both adjacent tubes are not active, the next-to-adjacent tubes, 3 and 4, will be searched. With the attached hit in this layer, referring as H1, the tracklet is updated and extrapolated to the next layer, till the last layer of the inner axial tubes. 

The track finder starts to enter the stereo layers. At the transition from axial to stereo layer, a wider search window of $\pm$6 tubes is applied. Inside the stereo layer, a normal search and attach process will be performed with 2 adjacent and 2 next-to-adjacent tubes.

After finishing the stereo layers, the track finder enters the outer axial layers. As in the previous step, a wider search window of $\pm$6 tubes is applied at the transition from stereo layer to axial layer. Using this track finder, tracklet candidates consisting of a list of axial and stereo straws are formed.

The track finder is tolerant towards a missing layer. However, in the case that no hit is found in two consecutive layers, the track finder quits the current track finding. A track candidate is only delivered to the next step when there are at least 3  axial hits and 2 skewed hits.

\subsection{Track fitter}

With a group of hits in one tracklet obtained from the track finder, a track fitter is designed to extract the helix parameters of the track. Here, a least squares fit is used based on minimising the mean square distance from the fitting curve to the data points.

The axial hits are used for the measurement of the helix curvature related to the transverse momentum. In the XY plane, a helix is projected to a circle expressed in equation~\ref{eq_circle} with 3 parameters: a, b, c. In the current implementation, parameter c is assumed as zero. As a consequence, only tracks from the primary vertex are properly reconstructed. As for particles from secondary vertices, the inclusion of MVD hits is required which is not considered in the present version of the algorithm.

\begin{equation}\label{eq_circle}
x^2 + y^2 + ax + by + c = 0
\end{equation}

\begin{equation}\label{eq_xy}
\Delta^2 = \Sigma(x_i^2 + y_i^2 + ax_i + by_i + c)^2 (1/d_i)^2
\end{equation}

\begin{equation}\label{eq_solution}
a = \frac{S_{yy}(-S_{xxx}-S_{xyy}) - S_{xy} (-S_{xxy}-S_{yyy})}{S_{xx}S_{yy} - S_{xy}S_{xy}}
\end{equation}
\begin{equation}
b = \frac{-S_{xy}(-S_{xxx}-S_{xyy}) + S_{xx}(-S_{xxy}-S_{yyy})}{S_{xx}S_{yy} - S_{xy}S_{xy}}
\end{equation}

Parameters a and b are determined by minimizing $\Delta^2$ in equation~\ref{eq_xy}. Here, the index i indicates the $i_{th}$ hit. $x_i$ and $y_i$ are the wire positions of the hit, $d_i$ is the drift radius of the hit.
$S_{xx} = \Sigma(x_i x_i)$, $S_{xy} = \Sigma(x_iy_i)$, $S_{yy} = \Sigma(y_iy_i)$, $S_{xxx} = \Sigma(x_ix_ix_i)$, $S_{xxy} = \Sigma(x_ix_iy_i)$, $S_{xyy} = \Sigma(x_iy_iy_i)$, $S_{yyy} = \Sigma(y_i y_iy_i)$.

Rather than the coordinates of the hit in three dimensions, each measurement of a straw tube provides a drift circle. A fit fully considering drift circles by minimising the mean square distance from the fitting curve to drift circles, would make the calculation too complicated and not suitable for implementation in VHDL. Thus, a simplification by using the central position of each drift circle with the reversed drift radius as a weight is performed, as shown in  equation~\ref{eq_xy}. The parameters a, b are determined in equation~\ref{eq_solution}. By this method, a 6\% transverse momentum resolution is achieved at 1~GeV/c. 

To improve the resolution, a second iteration fit is applied. Using the extracted momentum from the previous fit, the relation between each drift circle and the track can be determined. The center of a drift circle is either inside or outside of the helix circle in  equation~\ref{eq_circle}. With this information, a point on each drift circle is chosen, as the best conjecture of the trajectory. Using these points selected at drift circles, the second iteration fit improves the resolution to 3\% at 1 GeV/c. No significant improvement of resolution was observed with further iterations.

After this stage, three of the five parameters of the track helix are known (the radius R, the position of the helix center in the X-Y plane). The remaining two parameters of the helix are determined by using the hits of the skewed straws.

\begin{figure}[ht!]
\centering
\includegraphics[width=3.5in]{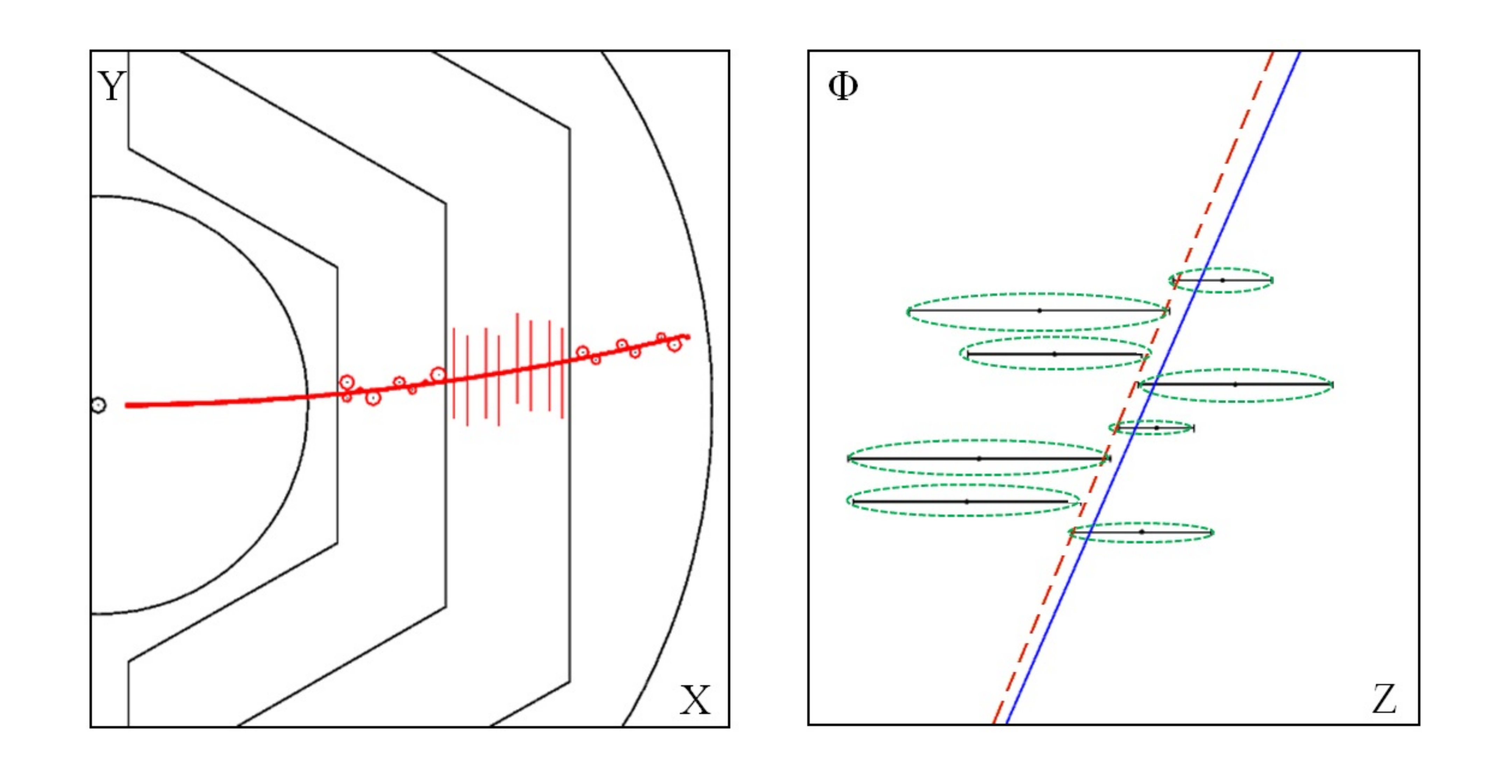}
\caption{Example of a single track displayed in X-Y view (left) and $\phi$-Z view (right). In the $\phi$-Z view, the blue solid line represents the fit result from the first iteration using central position of each hit. The red dashed line shows the fit result after the second iteration. \label{track_pt_pz}}
\end{figure}

The drift isochrone of a skewed straw intersects the cylinder on which the helix lies. This intersection represents approximately an ellipse~\cite{TDRSTT}, as illustrated in Fig.~\ref{track_pt_pz} (right). The helix trajectory is a straight line on the lateral surface of the helix cylinder ($\phi = KZ + \phi_0$ in $\phi$-Z plane) and tangent to the ellipses of the skewed straws, as explained in reference ~\cite{TDRSTT}. K and $\phi_0$ are the remaining two parameters of the helix. As for the determination of the transverse momentum, a least squares fit is used here by minimizing the mean square distance in equation~\ref{eq_phi_z}. Again, the fit is designed to run with two iterations, with the first iteration to solve the left-right ambiguity and the second iteration to improve the resolution.

In Fig.~\ref{track_pt_pz}, a single track is displayed. The left plot shows the X-Y view of the track. Here, the small red circles of different size indicate the drift circles. The vertical red lines indicates the projection of skewed tubes. The right plot displays the $\Phi$-Z plane. Here, the horizontal line segment inside an ellipse represents the major axis. A first iteration fit using the central positions of the horizontal line segments results in the blue solid line. The second iteration with the left-right ambiguity  solved, improves the fit as shown by the red dashed line.

\begin{equation}\label{eq_phi_z}
\Delta^2 = \Sigma(\phi_i - KZ_i - \phi_0)^2 (1/d_i)^2
\end{equation}

Here, $\phi_i$ and $Z_i$ are the wire position of the hits in $\phi$-Z plane. K and $\phi_0$ are two parameters of the helix.

\section{VHDL implementation}

\begin{figure}[ht!]
\centering
\includegraphics[width=3.5in]{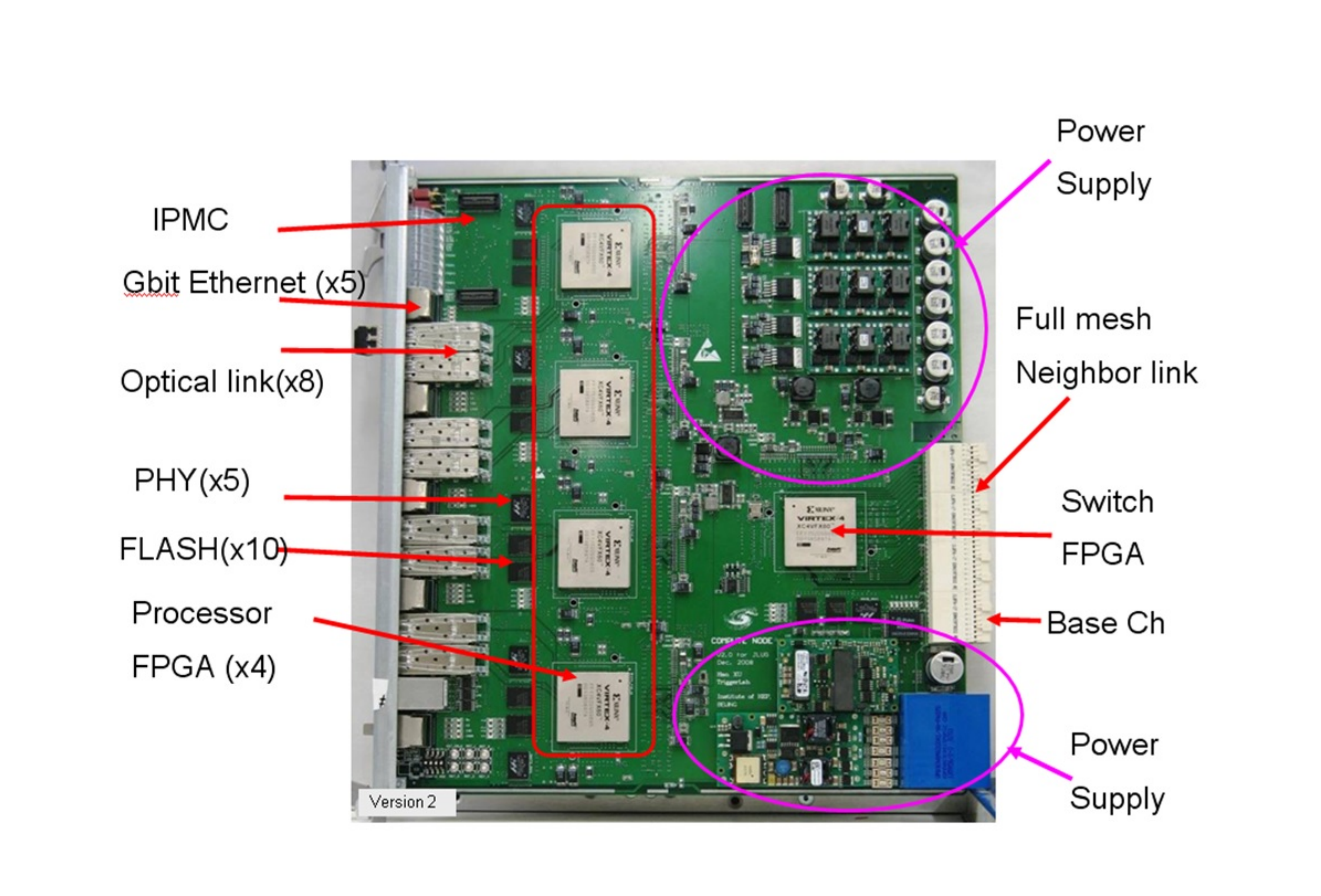}
\caption{The Compute Node used in this study, equipped with 5 Virtex 4 FX60 FPGA chips. \label{test_bench}}
\end{figure}

To test the tracking algorithm in the FPGA chip, an experimental test bench was set up, with one PC connected to an Advanced Telecom Computing Architecture (ATCA) based compute node (CN) \cite{CN} via a GBit Ethernet link. In Fig.\ref{test_bench}, a CN and the is shown. The PC is emulating the detector by sending straw tube hit information from Monte Carlo simulations to the tracking algorithm. The CN which is designed as a generic solution for trigger and data acquisition was equipped with 5 Virtex 4 FX60 FPGA chips. It is plugged into an ATCA shelf, capable of supporting
up to 14 CNs. The features of CN can be found in \cite{CN}.

The algorithm has been implemented in VHDL. In the code, fixed-point numbers are used. The hit information, such as the position x y z of a wire, is expressed as  24 bit fixed-point numbers, with the first bit as a sign bit the following 7 bits as the integer part and the last 16 bits as the fraction. Each tube has its own identification number, with 3 bits for the segment number, 5 bits for the layer number and 6 bits for the tube number.

\begin{figure}[ht!]
\centering
\includegraphics[width=3.5in]{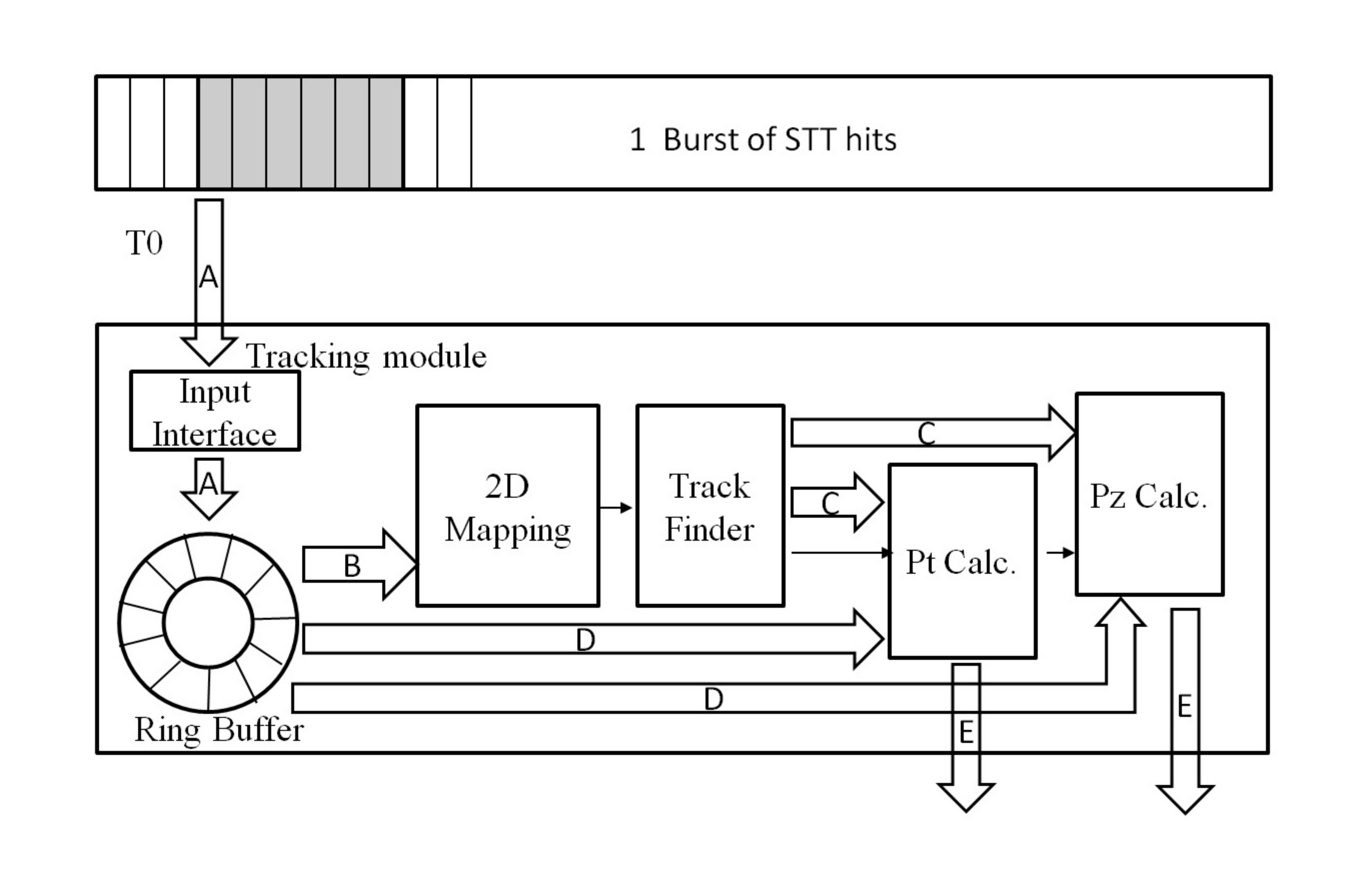}
\caption{Simplified block-diagram of the algorithm in VHDL. \label{schematic_vhdl}}
\end{figure}

The simplified block diagram of the implementation is shown in Fig.~\ref{schematic_vhdl}. The data sources for the algorithm include a collection of STT hits in a burst and a list of event start times T0. 


Following the data flow, the algorithm is comprised of:

1) The Input Interface module taking charge of reading the STT hits. Hits in the window of $T_0$ to $T_0+200$ ns are packed as one burst event which is copied to a ring buffer.

2) The ring buffer is a circular FIFO with a depth of 1024 and a width of 96 bits, composed of 72 bits for position x y and z, 14 bits for the ID of the tube and 10 bits for the arrival time of the hit in the burst. 

3) The 2D mapping module contains a dual port block RAM, with a depth of 16384 and width of 16 bits. Each tube has one correspondence in the 2D map. Reading one hit, its corresponding bin in the 2D map is marked as "Occupied". At the last hit, the 2D mapping task is finished and a start signal is passed to the next module, the Track Finder.

4) Once a start signal is received, the Track Finder starts to work. Hits from the innermost layer are saved in one array and treated as seed. Starting from each seed, the Track Finder searches its neighbours in next layer, attaches it if it is "Occupied". When one tracklet is completed, a start signal is sent to the next module, the Pt Calc.

5) The Pt Calc module reads the hit index from each tracklet and calculates the transverse momentum. There are two input data streams, one from the Track Finder with only the index of hits in a tracklet, and another from the ring buffer containing the full information of corresponding hits. 11 Xilinx IP (Intellectual property) cores for 32 bit multiplication are used, which have a latency of 6 clock cycles. At the end of the first iteration, the module updates the position of each hit by choosing one point around the drift circle and starts the second iteration. When the second iteration is finished, a signal is sent to the Pz Calc module.

6) The Pz Calc is designed to calculate the Pz information in two iterations. The transverse momentum from the previous module is read in to calculate the intersection with  skewed straws.

\begin{figure}[ht!]
\centering
\includegraphics[width=3.5in]{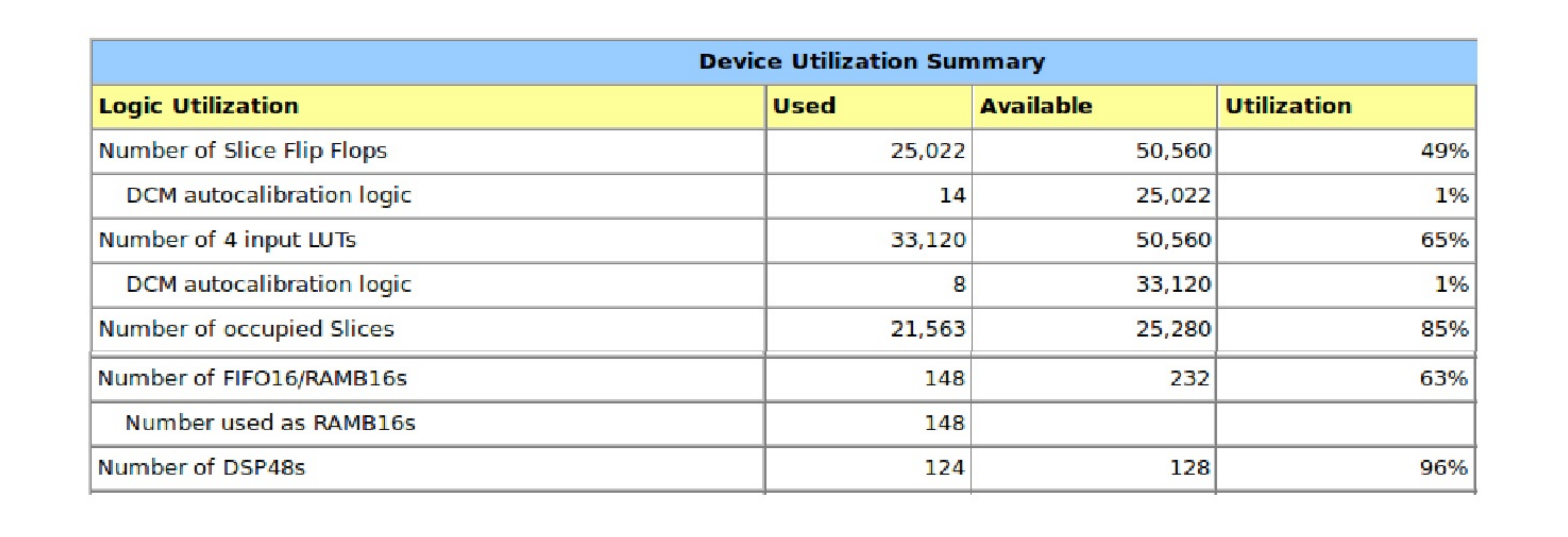}
\caption{Device utilization summary for the design. \label{device_util}}
\end{figure}

These modules are optimised to operate in a pipelined mode. The first 3 modules, the input interface, the ring buffer and 2D mapping modules work hit-based, with one clock cycle required for one hit. In contrast, the last 3 modules, the track finder, the Pt Calc and Pz Calc modules work tracklet-based. Once a tracklet is found by the Track Finder, the following Pt Calc module starts immediately while the Track Finder continues with the search for the next tracklet. This reduces the execution time. About 700 clock cycles (7 $\mu$s) are required for one event with 6 charged track (about 100 hits). The device utilisation summary table for the design is shown in Fig.~\ref{device_util}.

\section{Performance results}

\begin{figure}[ht!]
\centering
\includegraphics[width=3in]{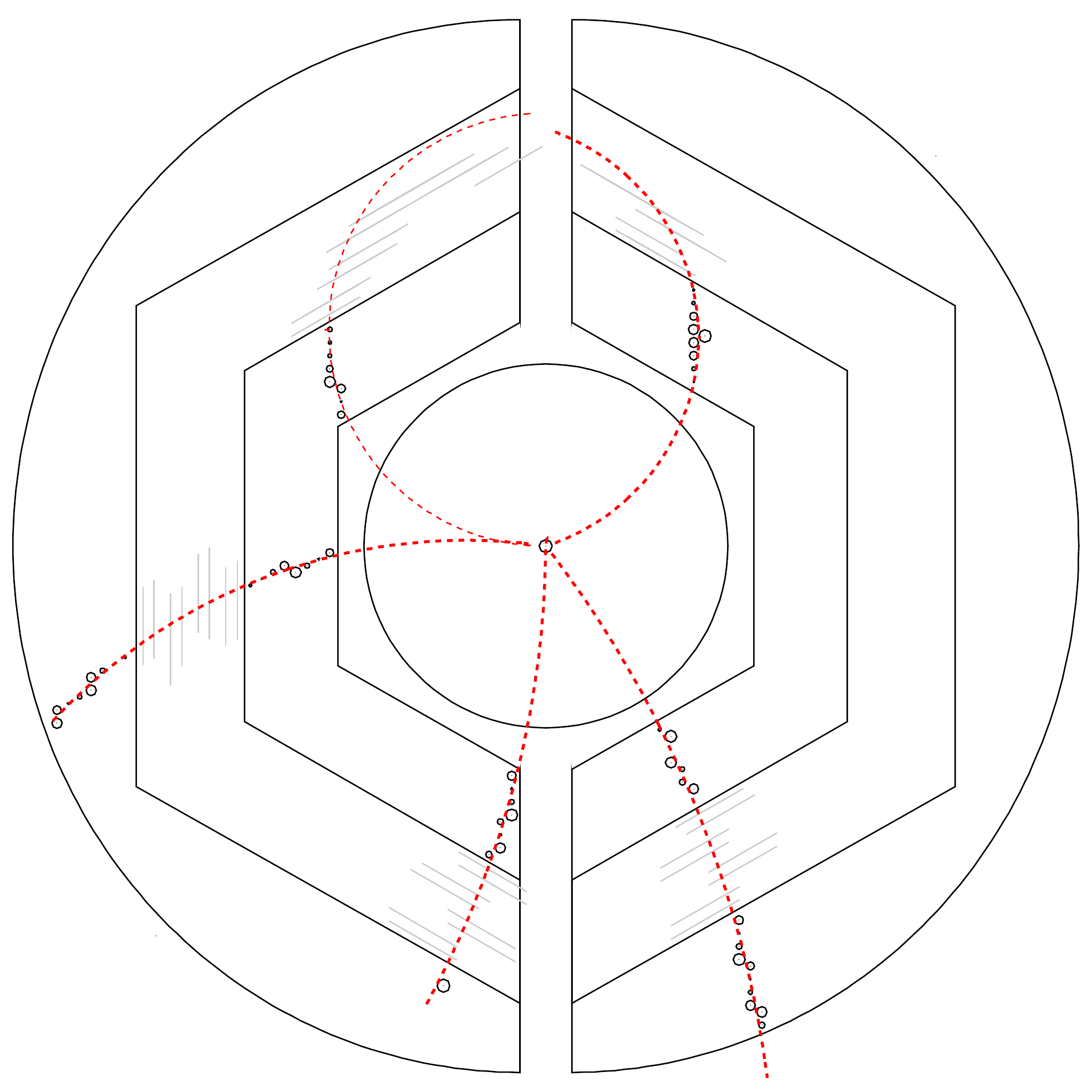}
\includegraphics[width=3in]{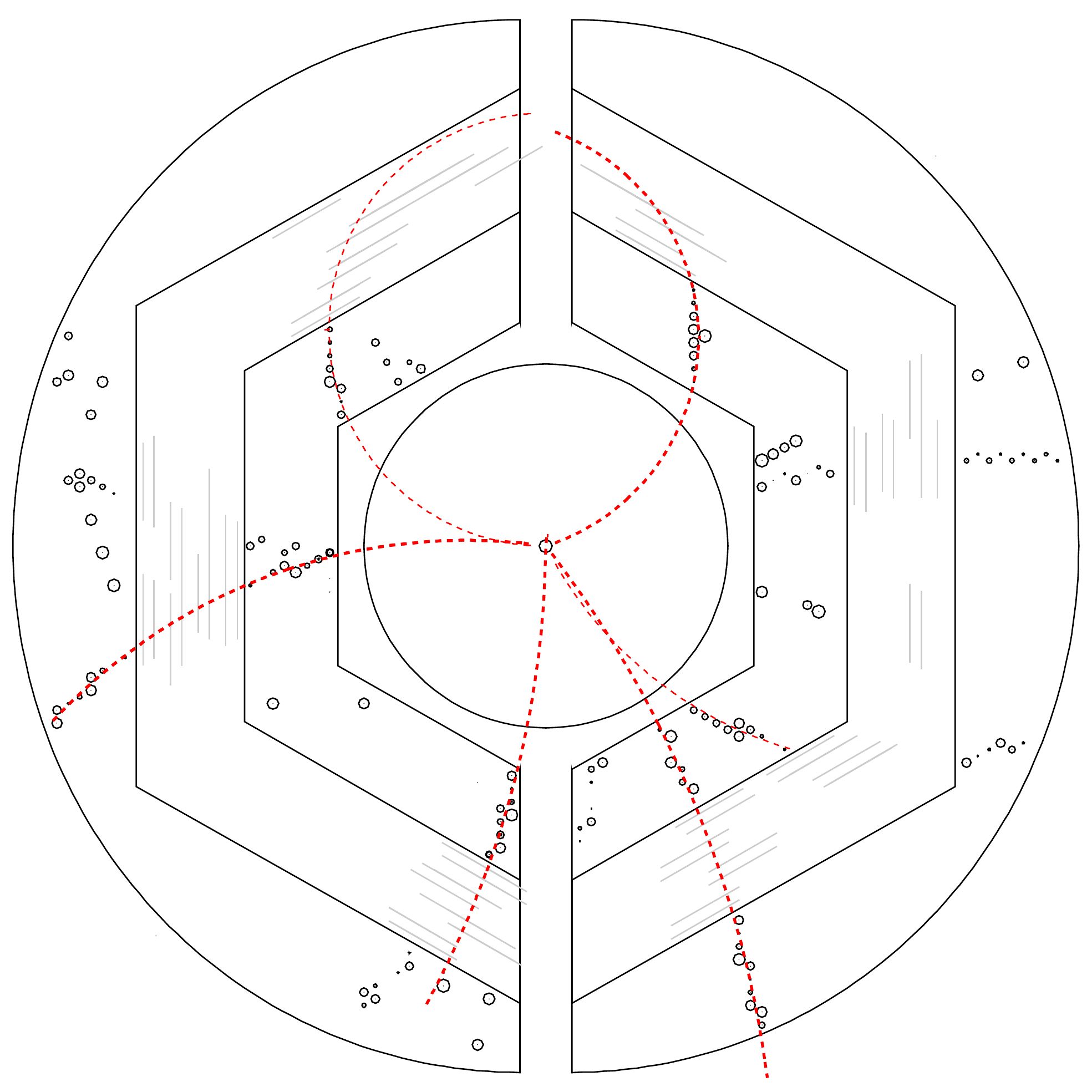}
\caption{Example of one burst event at low event rate (top) and high event rate of 20 MHz (bottom). The red circles represent the tracks found by the algorithm. \label{dpm_event_at_diff_rate}}
\end{figure}

In the investigation of the performance of the algorithm, we first run the algorithm with well separated events, corresponding to very low event rate. Then tests at different event rates are performed. An example of one burst event at low or high event rate is shown in Fig.~\ref{dpm_event_at_diff_rate}.

\subsection{Without overlapping events}

At very low event rate, events are well separated, with no overlapping. Event-based simulation, which does not consider overlapping events is used to generate Monte Carlo (MC) samples in the framework of PandaRoot \cite{pandaroot}.

The tracking efficiency is defined as $\epsilon = N_{rec}/N_{gen}$, where $N_{rec}$ is the number of tracks reconstructed and $N_{gen}$ is the number of generated tracks which has at least 3 hits in the STT. A reconstructed track is required to be within 6 standard deviations with respect to the momentum of the simulated one. 

\begin{figure}[ht!]
\centering
\includegraphics[width=3.5in]{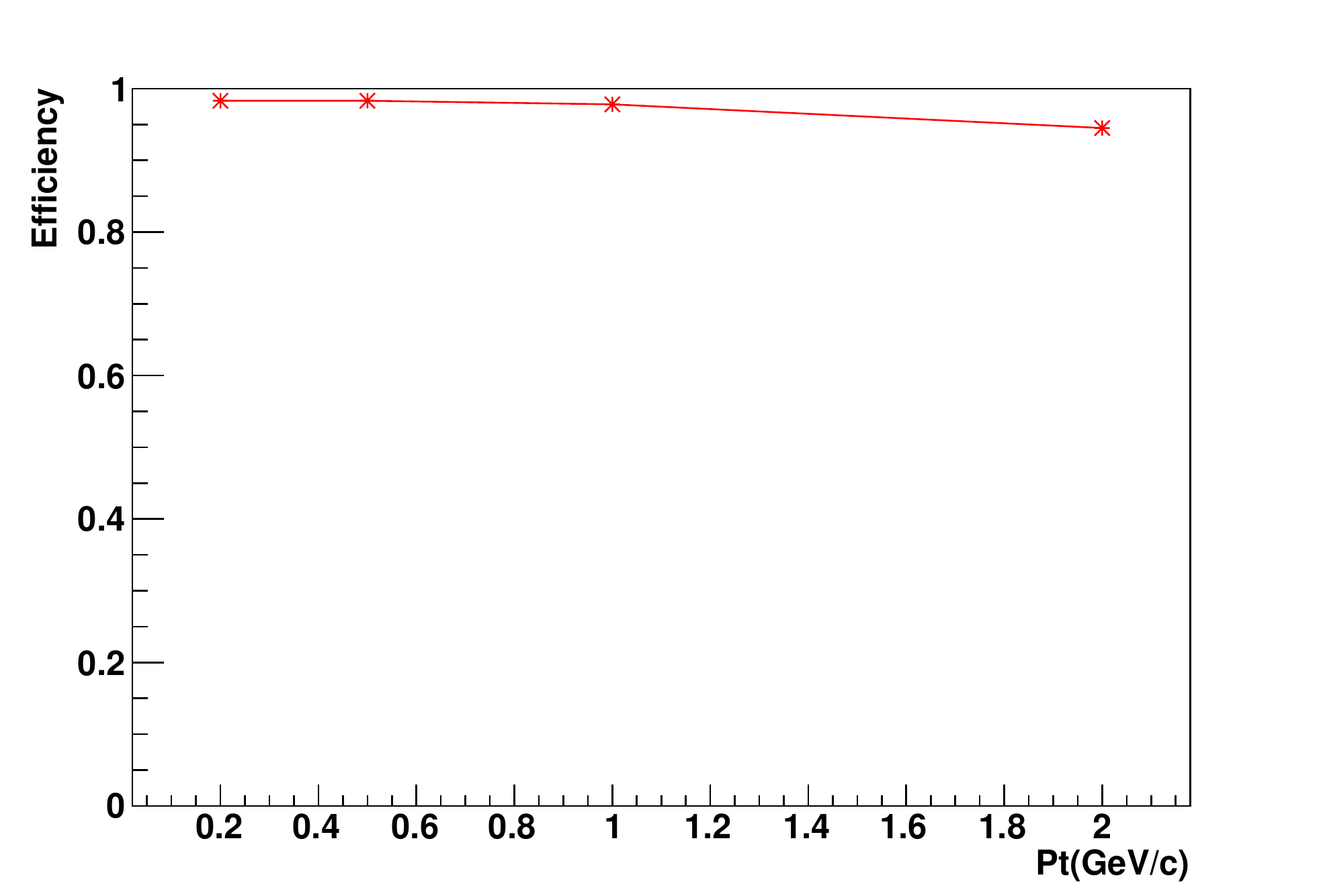}
\includegraphics[width=3.5in]{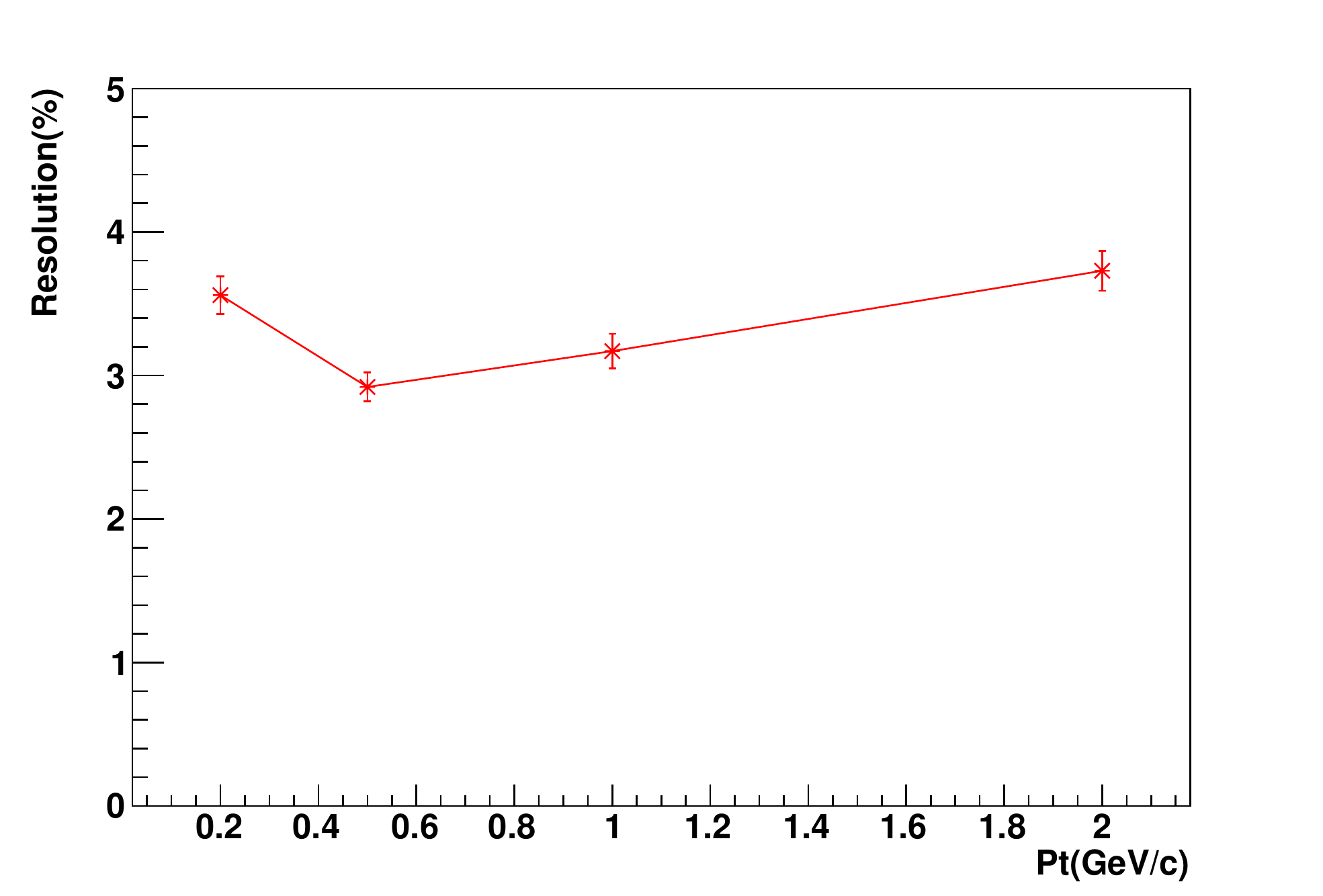}
\caption{The tracking efficiency and the resolution as a function of the transverse momentum. \label{effi}}
\end{figure}

The tracking efficiency is studied at four transverse momentum bins, as shown in Fig.\ref{effi}. At high transverse momentum (with $p_t \geq$ 2 GeV/c), a drop of efficiency is observed. This is due to the reduced bending of tracks at high transverse momentum. A track, appearing like a straight line originating from the interaction point has a larger probability to trigger tubes which are at the same side of its trajectory, which makes the momentum extraction to be more difficult. The resolution $\sigma(p_t)/p_t$  at different momentum bins is shown in Fig.\ref{effi}. The resolution $\sigma(p_z)/p_z$ depends on $p_t$ because $p_z$ is calculated with the input of $p_t$. In the case of $p_t$ = 1 GeV/c, $\sigma(p_z)/p_z$ at 0.5 GeV/c (1.0 GeV/c, 2.0 GeV/c) is 3.1\% (3.8\%, 4.2\%).


The hit information from the prepared MC samples is transmitted into the connected FPGA chip, in which the track finder and track fitter are run accordingly as explained in the previous section. The resulting momentum information from the algorithm is send back to the PC. The left plot in Fig.\ref{dpm_event_at_diff_rate} displays one example event that was analysed by the FPGA. When increasing the interaction rate (to 20 MHz), the average time interval between two events becomes smaller (to 50 ns), which will cause the overlap of events due to the potentially longer drift times. This behaviour can be seen by observing the same event in  Fig.\ref{dpm_event_at_diff_rate} (bottom). With an interaction rate of 20 MHz, hits from adjacent events are mixing up.

\subsection{With overlapping events}

The overlapping of events is simulated in a time-based simulation, in which hits are arranged in bursts by its time stamp information. To study the impact of the event rate, MC samples with an average of three charged tracks per event are generated with time-base simulation at four different event rates : no overlapping (referred as 0 MHz), 2 MHz, 10 MHz, and 20 MHz.
The tracking efficiencies at different event rates are shown in the left plot of Fig.~\ref{event_rate}. A drop of efficiency at high event rate is observed.

With increasing event rate, the number of fake tracks increases, as seen in Fig.~\ref{event_rate} (bottom). { Fake tracks arise from overlapping events and false combinations of track segments from the inner and outer axial layers (combinatorial fake tracks). The impact of these two sources of fake tracks are studied separately. Using generated DPM (Dual Parton Model) ~\cite{dpm} events representing the bulk annihilation coss section, the number of tracklets in one event at 20 MHz interaction rate reaches about 26, comparing to about 5 tracklets per event at low event rate (event-based simulation). In order to study combinatorial fake tracks, MC samples with different numbers of charged tracks per event are generated. 

From 3 tracks to 8 tracks per event, the number of combinatorial fake tracklets has no obvious increase. This is due to a the matching window for inner and out layers. As long as the track multiplicity is not exceedingly high, as shown in ~\cite{TDRSTT}, the number of combinatorial fake tracks  will not increase dramatically.


\begin{figure}[ht!]
\centering
\includegraphics[width=3.5in]{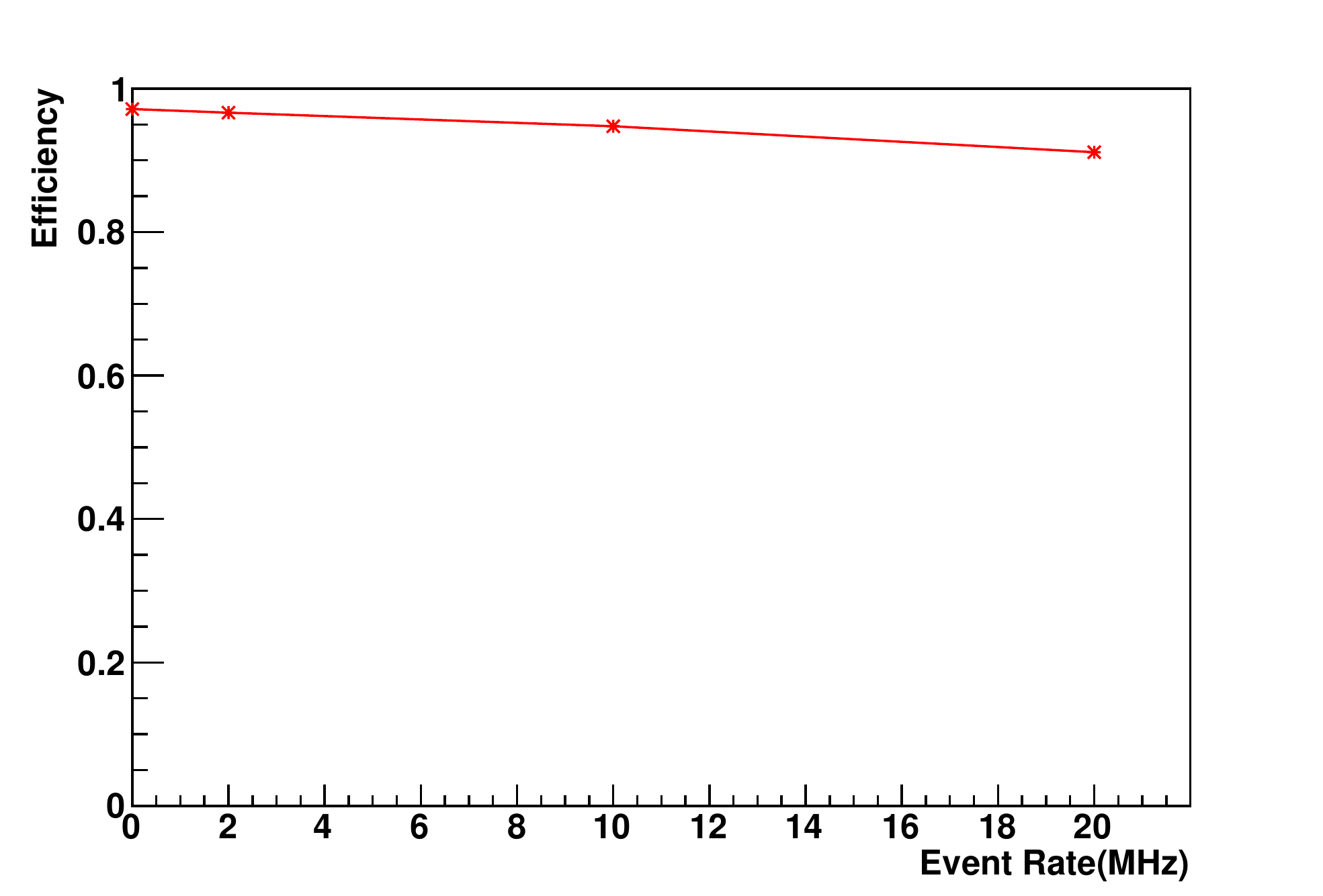}
\includegraphics[width=3.5in]{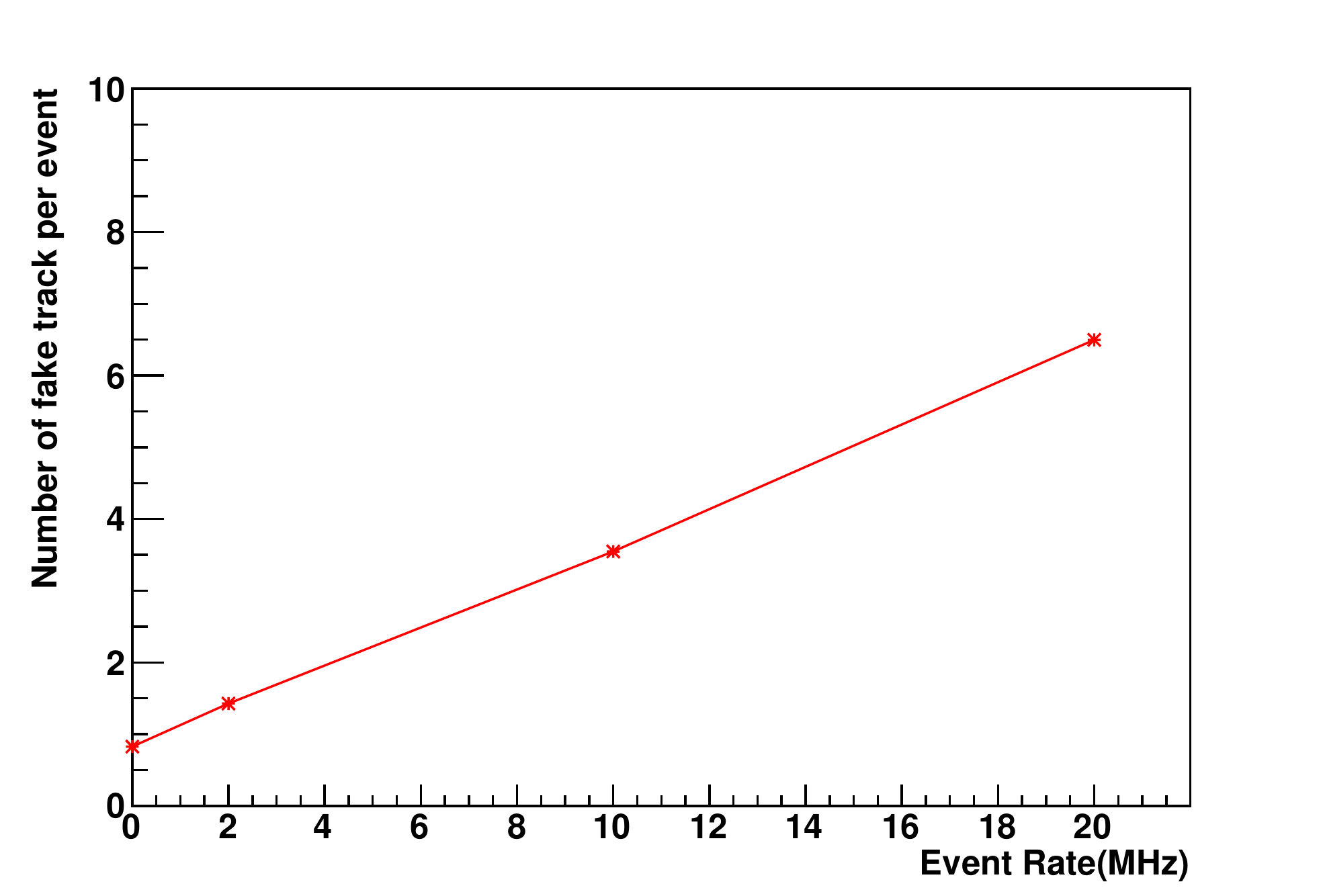}
\caption{Tracking efficiency (top) and  number of fake tracks (bottom) as a function of the event rate.
 \label{event_rate}}
\end{figure}

Considerung an event with correct $T_0$ assignment, hits from overlapping events will be analysed with the wrong $T_0$ and will have drift radii that are either too small or too large, as shown in Fig.~\ref{schematic_t0_extraction}.  This feature can be used to disentangle overlapping events, based on the track quality parameter.

As online tracking algorithm, the latency is an important parameter. The timing analysis and tests show a consistent latency of 7 $\mu$s for one event with 6 tracklets. 

\section{Summary and outlook}
 
A FPGA-based tracking algorithm has been designed for the data acquisition system of the PANDA experiment and tested on a Virtex4 FX60 FPGA. The algorithm features  high tracking efficiency, sufficient momentum resolution (3\% (4\%) for $p_t$ ($p_z$) at 1GeV/c) and low latency (7 $\mu$s to process one event with 6 tracklets). Comparing to a CPU chip (single core Intel Xeon E5520 at 2.26 GHz), where the algorithm was implemented in C++,  about 40 ms are required to process one event. An improvement of more than a factor of 5000 for  the specific FPGA implementation  compared the CPU implementation is obtained.

\begin{figure}[ht!]
\centering
\includegraphics[width=3.5in]{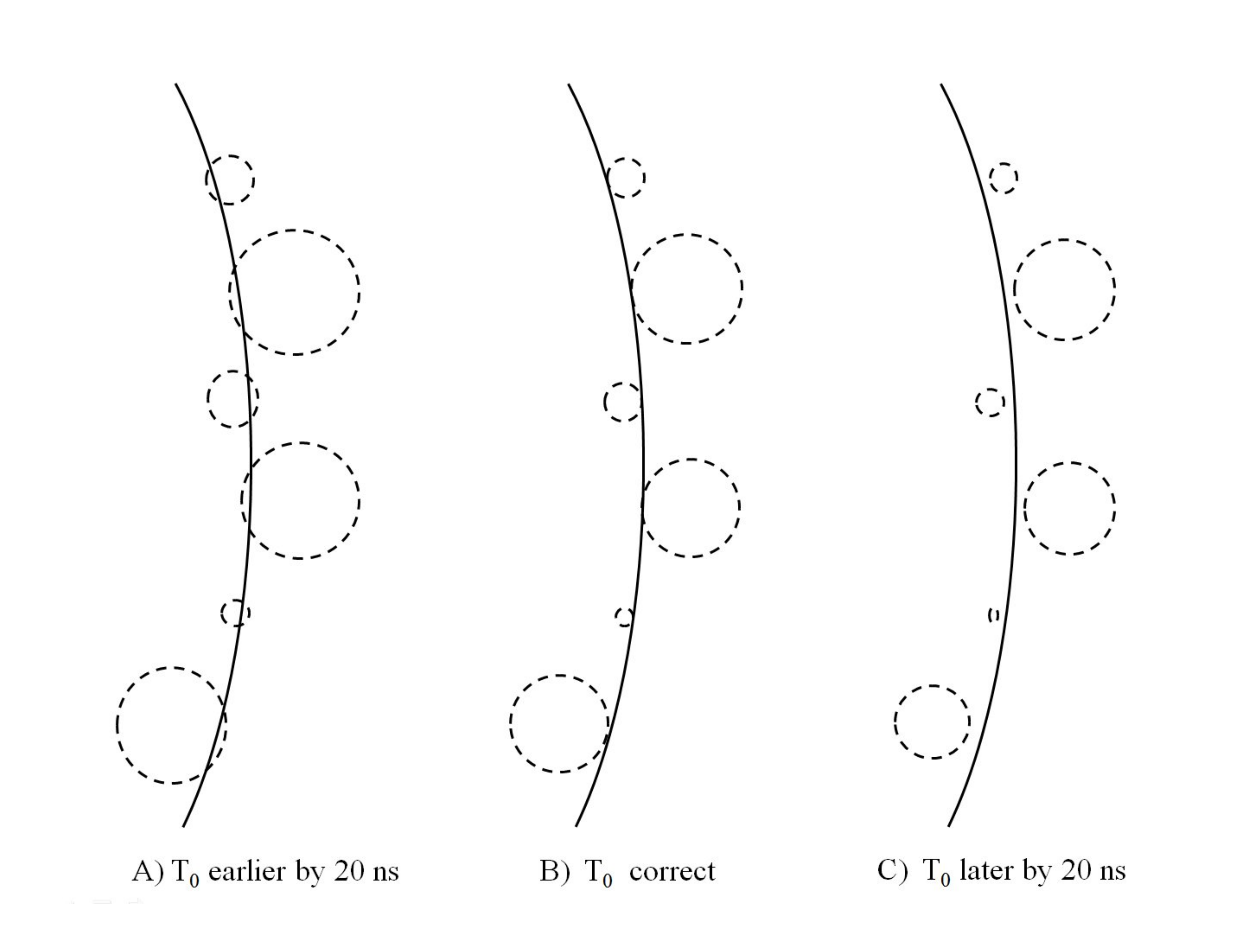}
\caption{$T_0$ assignment. Hits from overlapping events will be analysed with the wrong $T_0$ and will have drift radii that are either too small or too large, resulting in bad track quality. \label{schematic_t0_extraction}}
\end{figure}



Using only the information from the STT, the algorithm is now optimized for primary particles. As for particles from secondary vertices, the inclusion of MVD hits is needed, which is subject to future work. 

The algorithm assumes $T_0$ being provided by external fast detectors. Track quality can be used to discriminate tracks from overlapping events with different $T_0$.

 However, even in the absence of external $T_0$ information, the algorithm can be extended to derive $T_0$ based on the STT information only. This is illustrated in Fig.~\ref{schematic_t0_extraction}. Here, as discussed above, wrongly assigned $T_0$s result in either too large or too small drift circles and hence bad track quality. The closest distance of the track to the drift circles can be used to obtain the correct $T_0$. Based on this method,  tracking and self-consistent $T_0$ determination can be performed simultaneously, which is foreseen as a future extension of the algorithm.




%

\end{document}